\documentclass[12pt]{article} 
\usepackage{epsf}
\usepackage{pazha}
\tightenlines
\def\phflux{phot cm$^{-2}$ s$^{-1}$ keV$^{-1}$}
\def\a{$^{\mbox{\small a}}$}
\def\b{$^{\mbox{\small b}}$}
\def\c{$^{\mbox{\small c}}$}
\def\d{$^{\mbox{\small d}}$}
\def\型{$\pm$}
\def\ergs{$\mbox{erg s}^{-1}$}

\begin{document}

{\it To be published in ``Astronomy Letters'', 2001, v.27, p.297} 

\bigskip

\title{\bf The X-ray Source SLX 1732-304 in the Globular Cluster Terzan~1:
  The Spectral States and an X-ray Burst}  

\author{\bf M.N.Pavlinsky\affilmark{1}, S.A.Grebenev\affilmark{1},
  A.A.Lutovinov\affilmark{1},\\  
 R.A.Sunyaev\affilmark{1,2} and A.V.Finoguenov\affilmark{1}}   

\affil{
{\it Space Research Institute, Russian Academy of Sciences, Profsoyuznaya
str., 84/32, Moscow, 117810 Russia}$^1$\\
{\it Max-Planck Institut f{\"u}r Astrophysik, Garching, Germany}$^2$}

\vspace{2mm}
\received{29 June 2000}

\sloppypar 
\vspace{2mm}
\noindent

The results of ART-P/Granat observations of the X-ray burster SLX1732-304 in
the globular cluster Terzan 1 are presented. The X-ray (3-20 keV) fluxes
from the source differed by more than a factor of 4 during the observing
sessions on September 8 ($F_{\rm X}\simeq6.95\times10^{-10}$ erg cm$^{-2}$
s$^{-1}$) and October 6, 1990 ($F_{\rm X}\simeq1.64\times10^{-10}$ erg
cm$^{-2}$ s$^{-1}$). The intensity variations of SLX1732-304 were
apparently accompanied by variations in its hardness: whereas the
source in its high state had the spectrum with a distinct exponential
cutoff typical of bright low-mass X-ray binaries, its low-state
spectrum could be satisfactorily described by a simple power law with
a photon index $\alpha\simeq1.7$. During the ART-P observation on
September 8, a type I X-ray burst was detected from SLX1732-304. \\

\clearpage
 
\section*{INTRODUCTION}

\vskip -5pt 

Persistent X-ray emission from the globular cluster Terzan 1 was first
discovered in 1985 with the XRT telescope onboard the SPACELAB-2 space
laboratory (Skinner et al., 1987). A flux of $1.84\times10^{-10}$ erg
cm$^{-2}$ s$^{-1}$ was detected in the 3-30 keV energy band. The XRT
telescope was one of the first coded-aperture X-ray instruments that allowed
the sky images within the field of view to be reconstructed with an angular
resolution of several arcminutes. The localization with XRT showed that the
new X-ray source, designated as SLX1732-304, closely coincided in position with
the cluster center. Since two X-ray bursts had been previously detected from
this region with the HAKUCHO instruments (Makishima et al., 1981), Skinner
et al. (1987) concluded that the discovered source was an X-ray burster,
i.e., a neutron star with a weak magnetic field in a low-mass binary
system. Weak \mbox{X-ray} emission was detected from the source with
approximately the same flux during its subsequent EXOSAT (Parmar et
al., 1989), Granat (Pavlinsky et al., 1995; Borrel et al., 1996),
ROSAT (Johnston et al., 1995), and RXTE (Guainazzi et al., 1999;
Molkov et al., 2001) observations. Only the recent BeppoSAX
observations (Guainazzi et al., 1999), when the 2-10 keV flux from the
source was a factor of $\sim 300$ lower than its normal value,
constitute an exception. Of particular interest are also the
ART-P/Granat observations (Pavlinsky et al., 1995), when the source
was observed in states with different intensities and when an X-ray
burst was detected from it. 

Note that the distance to the globular cluster Terzan 1 has been known
poorly until recently. In several studies (Skinner et al., 1987; Pavlinsky
et al., 1995; Borrel et al., 1996), it was assumed to be $d\sim 10$
kpc. Optical observations suggested a more modest value, $d\simeq 5.9$ kpc
(see Johnston et al., 1995 and references therein). In general, this
estimate is confirmed by recent observations of the cluster with the ESO
telescopes, $d\simeq 4-5.5$ kpc (Ortolani et al., 1993), and with the Hubble
Space Telescope, $d\simeq 5.2\pm0.5$ kpc (Ortolani et al., 1999). In our
subsequent analysis, we use $d=5.2$ kpc.

In this paper, we present the results of our detailed timing and spectral
analyses of the ART-P/Granat observations for the X-ray source in Terzan 1
obtained in 1990-1992.\\

\section*{OBSERVATIONS}

The ART-P telescope onboard the Granat orbiting astrophysical observatory
consists of four independent modules; each module is a position-sensitive
detector with a geometric area of 625 cm$^{2}$, equipped with a collimator
and a coded mask whose opaque elements always cover half of the
detector. The instrument has a 3\fdg4$\times$3\fdg6 field of view (FWHM) and
an angular resolution of 5 arcmin (the element size of the coded
mask). Because of the detector's higher spatial resolution ($\sim 1.25$
arcmin), the localization accuracy for discrete sources is several-fold
higher. The telescope is sensitive to photons in the energy range $2.5-60$
keV (the photon arrival time is accurate to within $\sim3.9$ ms, and the
ART-P dead time is 580 $\mu$s) and has an energy resolution of $\sim 22$\%
in the 5.9-keV iron line. See Sunyaev et al. (1990) for a more detailed
description of the telescope.

\begin{table}[t]

\centering
{\bf Table 1. }{ART-P observations of SLX1732-304 in 1990-1992.}\\

\vspace{5mm}

\begin{tabular}{r|r|c|c|c}
\hline
\hline
Date(UT) & Exposure, s & Efficiency\a,\% & Flux\b, mCrab& ${\rm
  N}_{\rm B}$\d \\ \hline
 8.09.90 &26680  &       42.0    & $25.0\pm1.0$&1\\
 9.09.90 &7370   &        2.5    & 107\c       &0\\
 6.10.90 &40460  &       29.1    & $7.4\pm1.6$ &0\\
 9.10.90 &6590   &        4.0    & 72.4\c      &0\\
23.02.91 &3040   &        2.8    & 460\c       &1\\
30.08.91 &4995   &        3.2    & 300\c       &0\\
19.09.91 &16950  &       20.9    & 25.8\c      &0\\
 2.10.91 &13970  &       13.3    & 39.6\c      &0\\
15.10.91&7400    &        1.8    & 239\c       &1\\
17.10.91 &6040   &        4.5    & 219\c       &0\\
18.10.91 &3960   &        3.3    & 325\c       &0\\
19.10.91 &7920   &        4.1    & 188\c       &3\\
19.02.92 &24530  &       14.4    & 14.3\c      &0\\
29.02.92 &5360   &      3.0     & 439\c        &1\\
 1.03.92 &4590   &      2.9     & 489\c        &0\\
 2.04.92 &7460   &      2.9     & 392\c        &0\\
 3.04.92 &5350   &      3.7     & 343\c        &0\\
 4.04.92 &6700   &      4.0     & 281\c        &0\\
 6.04.92 &6550   &      4.2     & 270\c        &0\\
 7.04.92 &3890   &      4.7     & 316\c        &1\\
 8.04.92 &1790   &      5.7     & 381\c        &0\\ \hline
\end{tabular}
\vspace{3mm}

\begin{tabular}{lll}
\a & Collimator transmission efficiency.\\
\b & Mean 6-20 keV flux.\\
\c & $3 \sigma$ upper limits.\\
\d & Number of bursts detected from sources within the field of view.\\ 
\end{tabular}
\end{table}

During regular observations of the Galactic-center region, SLX1732-304 was
within the ART-P field of view many times. The total exposure was
$6.6\times10^5$~s. In most cases, however, the source was at the very edge
of the field of view, in the region where the efficiency of observations was
low due to the drop in collimator transmission. Table 1 gives data only for
those sessions during which the mean efficiency exceeded 1\%. During a
session, the telescope axis could deviate from the pointing direction within
30 arcmin because of the spacecraft wobbling, and, accordingly, the
efficiency of observations could exceed its mean value during some
periods. We therefore selected time intervals for the sessions listed in the
table during which the instantaneous efficiency exceeded 1\% and used them
to determine the photon flux from the source and the actual exposure. Note
that the efficiencies given in the table also refer only to the selected
"good" parts of the sessions. The total exposure for these parts is
$2.2\times10^5$~s.

We see from Table 1 that in most sessions, no flux was detected from the
source, and only the corresponding upper limit was obtained. Nevertheless,
these sessions provide important information for the search for X-ray bursts
from the source and for estimating their recurrence time. Over the entire
period of observations, we detected only one burst that was unambiguously
identified with SLX1732-304. In addition, we detected nine more bursts that
we failed to localize and to identify with any source of persistent emission
in the Galactic-center region (Grebenev et al., 2001). Based on our data, we
can estimate the burst recurrence time, $11^{\rm h}.2<\tau<61^{\rm h}.1$
(the lower limit follows from the session duration on October 6, 1990, while
the upper limit follows from the total duration of observations,
$2.2\times10^5$ s). Since only three X-ray bursts were detected from this
source over more than twenty years of its observations, $\tau$ is most
likely closer to the upper limit.

A statistically significant persistent X-ray flux from the source was
detected by ART-P only during two sessions: September 8 and October 6, 1990
(see Table 1). During the former observation, the flux was a factor of $\sim
4$ higher than that during the latter observation. An X-ray burst was
detected from the source on September 8, 1990.\\

\section*{PERSISTENT SPECTRUM}

The X-ray spectrum of SLX1732-304 in quiescence can generally be described
by a power law with a photon index $\alpha\sim2.1$ modified at low energies
by interstellar absorption, which corresponds to a hydrogen column density
$N_{\rm H}\simeq 1.8\times10^{22}$ cm$^{-2}$ (Parmar et al., 1989; Johnston
et al., 1995; Guainazzi et al., 1999; Molkov et al., 2001). During the ART-P
observation on September 8, SLX1732-304 was virtually the only bright source
within the field of view (Fig. 1a), while the main target for the October 6
observation was the Galactic-center region (Fig. 2); apart from SLX1732-304,
there were such bright sources as A1742-294 and 1E1740.7-2942, as well as
the new sources GRS1734-292 and GRS1736-297 within the field of view
(Pavlinsky et al., 1992, 1994). The ART-P design allows the spectra of each
object within the field of view to be measured irrespective of their
number. The 3-20 keV fluxes measured on September 8 and October 6 were
$(6.95\pm0.18)\times10^{-10}$ erg cm$^{-2}$ s$^{-1}$ ($28.6\pm0.7$ mCrab)
and $(1.64\pm0.27)\times10^{-10}$ erg cm$^{-2}$ s$^{-1}$ ($6.7\pm1.1$
mCrab), respectively. Below, we therefore arbitrary call the source's states
during the September 8 and October 6 observations "high" and "low",
respectively. The changes in the spectrum occurred between these
observations are clearly seen in Fig. 3, which shows pulse-height spectra of
the source, and in Fig. 4, which shows the restored photon spectra. Best-fit
parameters for the high- and low-state spectra are listed in Table 2. The
interstellar absorption toward SLX1732-304 was fixed at $N_{\rm H}=
1.8\times10^{22}$ cm$^{-2}$ (Johnston et al., 1995). This value is in
satisfactory agreement with $N_{\rm H}\le 2.4\times10^{22}$ cm$^{-2}$ that
was directly determined from ART-P data (Pavlinsky et al., 1995). The
spectra were fitted by three models: a power law (PL), bremsstrahlung of an
optically thin plasma (BR), and Comptonization of low-frequency photons in a
cloud of high-temperature plasma (ST) suggested by Sunyaev and Titarchuk
(1980).

\begin{table}[t]

\centering
{\bf Table 2. }{Best-fit parameters for the spectrum of SLX1732-304}\\  

\vspace{5mm}

\begin{tabular}{r|c|c|c}
\hline
\hline
Date & Model\a  & Parameters &  $\chi^{2}_{\,N} (N)\,$\d\\ \hline
8.09.90 & {\bf ST}    & $I_{0}^{\mbox{\small b}}=0.255\pm0.069$ &  0.78\,(14)\\
        &             & $kT_e = 2.32\pm0.25$ keV & \\
        &             & $\tau =7.08\pm1.04$   & \\ [2mm]
        & {\bf BR}    & $I_{0}^{\mbox{\small b}} =0.217\pm0.012$ & 0.85\,(15)\\
        &             & $kT = 6.12\pm0.36$ keV & \\ [2mm]
        & {\bf PL}    & $I_{0}^{\mbox{\small b}} =0.765\pm0.087$ & 2.52\,(15)\\
        &             & $\alpha=2.55\pm0.62$   & \\[2mm] \hline
6.10.90 & {\bf ST}    & $I_{0}^{\mbox{\small b}}=0.017\pm0.024$  & 0.08\,(6) \\
        &             & $kT_e = 3.62\pm2.63$ keV& \\
        &             & $\tau=7.74\pm6.59$      & \\ [2mm]
        & {\bf BR}    & $I_{0}^{\mbox{\small b}} =0.019\pm0.004$ &0.09\,(7) \\ 
        &             & $kT = 26\pm26$ keV     & \\ [2mm]
        & {\bf PL}    & $I_{0}^{\mbox{\small b}}=0.029_{-0.016}^{+0.033}$ &
        0.13 \,(7) \\
        &             & $\alpha=1.66\pm0.38$   & \\[2mm]
        & {\bf ST}    & $I_{0}^{\mbox{\small b}}=0.106\pm0.004$ & 0.37\,(7)\\
        &             & $N_{\rm H}^{\mbox{\small c}}=31\pm17$\\
        &             & $kT_e=2.32$ keV & \\
        &             & $\tau=7.08$     & \\
 \hline
\end{tabular}
\vspace{3mm}

\begin{tabular}{lll}
\a & ST --& for the spectrum Comptonized in a disk with electron\\
   &      & temperature $kT_e$ and optical depth $\tau$,\\
   & BR --& for bremsstrahlung with temperature $kT$, \\
   & PL --& for a power-law spectrum with photon index $\alpha$.\\ 
\b & \multicolumn{2}{l}{$I_{0}$ is the flux at 1 keV (\phflux).} \\
\c & \multicolumn{2}{l}{$N_{\rm H}$ is the hydrogen column density
  ($\times10^{22}$ cm$^{-2}$).} \\ 
\d & \multicolumn{2}{l}{The $\chi^2$ value normalized to the number 
    of degrees of freedom $N$.}\\   

\end{tabular}
\end{table} 

As we see from Figs. 3 and 4 and from Table 2, the source's high-state 
spectrum exhibits a cutoff at high energies, resulting in a poor 
description of the spectrum by a pure power law. The ST and BR models 
describe the spectrum equally well; the bremsstrahlung fit yields the 
temperatures typical of many bright low-mass X-ray binaries, objects of 
the same type as that of SLX1732-304.

Since the source's low-state spectrum was measured with large statistical
errors, we cannot prefer one model to another. When this spectrum is fitted
by the ST and BR models, the temperatures that characterize the steepness of
the high-energy exponential cutoff are generally larger than those obtained
for the high-state spectrum. This is evidence that the source's hardness
increases with a transition to its low state. The best-fit parameters
of the low-state spectrum (see the low part of Table 2) for the Comptonization
model with fixed $kT_e$ and $\tau$ (the same as those obtained for the high
state) confirm this conclusion. The spectrum normalization $I_{0}$ and the
hydrogen column density $N_{\rm H}$, which characterizes the interstellar
absorption, were taken as the free parameters. This model simulates a
situation in which the transition between the states was associated with a
change in intrinsic intensity of the source's radiation (without any change
in spectral shape) or with the emergence of a scattering and absorbing cloud
of cold plasma on the line of sight. Although both these possibilities
cannot be completely rejected, we see that this model describes the
pulse-height spectrum worst. The long-term SIGMA/Granat observations of
SLX1732-304 (Pavlinsky et al., 1995; Borrel et al., 1996) also indicate
that the source has a hard Comptonized spectrum in its low state.

In Figs. 3 and 4, the solid and dashed lines represent the best fits to the
source's high- and low-state spectra with the Comptonization model.
According to the ART-P data, the source's mean 3-20 keV luminosity during
these states was $2.25\times10^{36}$ and $5.3\times10^{35}$~erg~s$^{-1}$,
respectively.\\

\section*{X-RAY BURST}

During the September 8, 1990 observation at $12^{h}16^{m}18^{s}$ (UT) ART-P
detected an intense X-ray burst (Fig. 5). Its localization showed that the
position of the burst source, R.A. =$17^{h}32^{m}34^{s}$,
Decl. =-30\deg29\arcmin26\arcsec (Pavlinsky et al., 1994), coincides, within
the error limits (120\arcsec at 90\% confidence), with the position of
SLX1732-304. Our result is illustrated by Fig.~1b, which shows an X-ray
(3-20 keV) map of the sky within the ART-P field of view during the first 26
s of the burst. It is of interest to compare it with a similar map obtained
during the entire observing session on September 8 (see Fig.~1a).

The time profile of the burst (an insert in Fig. 5) indicates that it
belongs to type I bursts, which are produced by a thermonuclear explosion on
the neutron-star surface and which are characterized by an abrupt rise
followed by a slow exponential decline in X-ray flux. A distinctive feature
of type I bursts is the so-called "cooling" of the radiation at the burst
exponential stage, i.e., a gradual softening of the spectrum and a decrease
in $e$-folding decline time with increasing energy. The burst time profiles
in Fig. 6 and the corresponding $e$-folding times of exponential decline
$t_{exp}$ show that the cooling state is well defined in this case.

Figures 4 (triangles) and 7 (filled and open circles) show three photon
spectra measured during the burst: the spectrum averaged over the entire
burst time ($\sim26$ s), the spectrum near peak flux (3-5 s after the burst
onset), and the spectrum at the end of the cooling stage (13-16 s). The
dashed, solid, and dotted lines represent the best fits to these spectra
with the blackbody model. Figure 7 clearly shows that the spectrum
appreciably softens by the end of the burst. To better trace the spectral
evolution, we singled out seven intervals in the burst profile and performed
a spectral analysis for each of them. The best-fit parameters for each
spectrum with the model of a blackbody sphere and the corresponding X-ray
(3-20 keV) flux are given in Table 3. The spectrum averaged over the entire
burst time can be described by the model of a blackbody sphere with
temperature $kT_{bb}\simeq1.83\pm0.17$ keV, radius $R_{bb}\simeq6.6\pm1.2$
km, and the 3-20 keV flux $F\simeq(1.92\pm0.19)\times10^{-8}$ erg cm$^{-2}$
s$^{-1}$, which corresponds to a luminosity $L_{\rm
X}\simeq6.2\times10^{37}$ \ergs.  Note that the blackbody model does not
satisfactorily describe the measured burst spectrum at hard ($h\nu>18$ keV)
energies (see Fig. 4).
\begin{table}[t]

\centering
{\bf Table 3. }{Spectral evolution of SLX 1732-304 during the X-ray burst}\\

\vspace{5mm}
\begin{tabular}{c|c|c|c|c}
\hline
\hline
Interval, s\a  & Flux, & $kT_{bb}$, & $R_{bb}$,  &$\chi^{2}_{\,N} (N)\,$\b\\ 
               & mCrab     & keV        &  km        &            \\ \hline

 ~0-2~       & ~904\型272 & 3.84\型1.17  & 1.88\型1.02  &0.43 (8) \\
 ~3-5~       & 2470\型340 & 2.49\型0.33  & 6.43\型1.59  &1.43 (8) \\
 ~6-8~       & 1549\型281 & 1.98\型0.33  & 7.80\型2.63  &0.66 (8) \\
 ~9-12       &  535\型236 & 1.77\型0.55  & 6.38\型4.21  &0.10 (5) \\
 13-16       &  578\型163 & 1.38\型0.40  & 9.76\型3.53  &0.84 (5) \\
 17-24       &  178\型86 &  1.00\型0.39  & 13.74\型14.74 &1.45 (8) \\

\hline
\end{tabular}
\vspace{3mm}

\begin{tabular}{ll}
\a & From the burst onset at $12^{h}16^{m}18^{s}$ (UT).\\
\b & The $\chi^2$ value normalized to the number of degrees of freedom
$N$.\\  

\end{tabular}
\end{table} 

Figure 8 shows the evolution of the source's bolometric luminosity, radius
$R_{bb}$, and temperature $kT_{bb}$ during the burst in the blackbody
model. The radius $R_{bb}$ is almost always smaller than the typical
neutron-star radius, $R_{ns}\simeq15$ km, suggesting that the thermonuclear
explosion responsible for the burst took place not instantaneously over the
entire neutron-star surface, but only in its part. However, a much more
probable reason is the oversimplified description of the measured
spectra. As was first shown by Sunyaev and Titarchuk (1986) and Ebisuzaki
and Nomoto (1986), the Comptonized spectra of the radiation emerging from
the photospheres of X-ray bursters during bursts can differ markedly from
Planck spectra. In particular, the Wien radiation component with a
temperature corresponding to the electron temperature of the outer
photosphere, which can be higher than the effective (blackbody) temperature,
dominates in the X-ray band.\\

\section*{CONCLUSIONS}

The ART-P observations of the X-ray burster SLX1732-304 in the globular 
cluster Terzan~1 are unique not only because the source was detected 
during them in states with different intensities, but also because, for 
the first time, it has become possible to carry out a detailed spectral 
study of the evolution of its radiation during an X-ray burst.

Our analysis shows that the transition between the high and low states was
apparently accompanied by a change in the source's hardness. In its high
state, the spectrum of the source was typical of bright low-mass X-ray
binaries whose compact object is a neutron star with a weak magnetic
field. It could be satisfactorily described by bremsstrahlung of an
optically thin thermal plasma with $kT\sim6$ keV or, equally well, by
Comptonization of low-frequency photons in a cloud of hot ($kT_e\sim2.3$
keV) electron plasma. In its low state, the source in the ART-P energy band
most likely had a power-law spectrum with no evidence of an obvious
high-energy cutoff. This means that the emission from the source in this
state originates in much hotter and more tenuous plasma.

The X-ray burst observed with ART-P on September 8 became the third burst
detected from this source over the entire period of its observations. Our
analysis shows that it is a classical type I burst; i.e., it is
characterized by an abrupt rise in flux followed by a slow exponential
decline, with the decline at hard energies ($t_{exp}\simeq2$ s) being faster
than that at soft energies ($t_{exp}\simeq11$ s). Our study of the source's
spectrum during the burst indicates that it can be satisfactorily described
by the model of a blackbody with a temperature smoothly falling during the
burst from 3.8 to 1.0 keV. The total energy release during the burst is
$E\simeq1.7\times10^{39}$ erg, which is equivalent to an explosion of
$M\simeq E/\epsilon_{N}\simeq 9.5\times10^{20}$ g of matter, where
$\epsilon_{N}\simeq0.002c^2$ is the helium burning efficiency. If we use an
estimate for the rate of accretion onto the neutron star in quiescence
($\dot M=L_{\rm X} R_{ns}/GM_{ns}$, where $L_{\rm X}$, $M_{ns}\simeq
1.4 M_{\sun}$ and $R_{ns}\simeq15$ km are the luminosity, mass, and radius
of the neutron star, respectively), then we can estimate the time it takes
for this amount of matter to be accreted onto the stellar surface, i.e., the
characteristic burst recurrence time: $\tau\simeq14^{\rm h}.8$ during the
high state and $\tau\simeq62^{\rm h}.7$ during the low state. When comparing
these estimates with the duration of the source's ART-P observations in its
high and low states (see Table 1), the burst detection in the former case
and its absence in the latter case seems natural. However, one should not
attach too great a significance to these estimates. Many bright X-ray
sources are known to show up as bursters only in a state with a certain
(usually moderately high) luminosity (Lewin et al., 1993; Molkov et al.,
1999). In this sense, the observation of the burst from SLX1732-304 in its
high state provides further evidence that this source is exceptional and
unique among X-ray bursters.

\section*{ACKNOWLEDGMENTS}

This study was supported by the Russian Foundation for Basic Research 
(project nos. 98-02-17056, 99-02-18178, and 00-15-99297). We wish to 
thank K.G. Sukhanov, flight director, the staffs of the Lavochkin 
Research and Production Center, RNIIKP, and the Deep Space 
Communications Center in Evpatoria, the Evpatoria team of the Space 
Research Institute (Russian Academy of Sciences), the team of I.D. 
Tserenin, and B.S. Novikov, S.V. Blagii, A.N. Bogomolov, V.I. Evgenov, 
N.G. Khavenson, and A.V. D'yachkov from the Space Research Institute who 
operated the Granat Observatory, provided the scientific planning of the 
mission, and performed a preliminary processing of telemetry data. We 
also wish to thank the team of M.N. Pavlinsky and the staff of the
former Research and Development Center of the Space Research Institute
in Bishkek who designed and manufactured the ART-P telescope. We wish
to thank V.Astakhov for the help in translating this paper in English. 

\pagebreak   

\pagebreak

\begin{figure}[t] 
\hspace{-0.5cm}{
\epsfxsize=170mm
\epsffile[20 320 590 620]{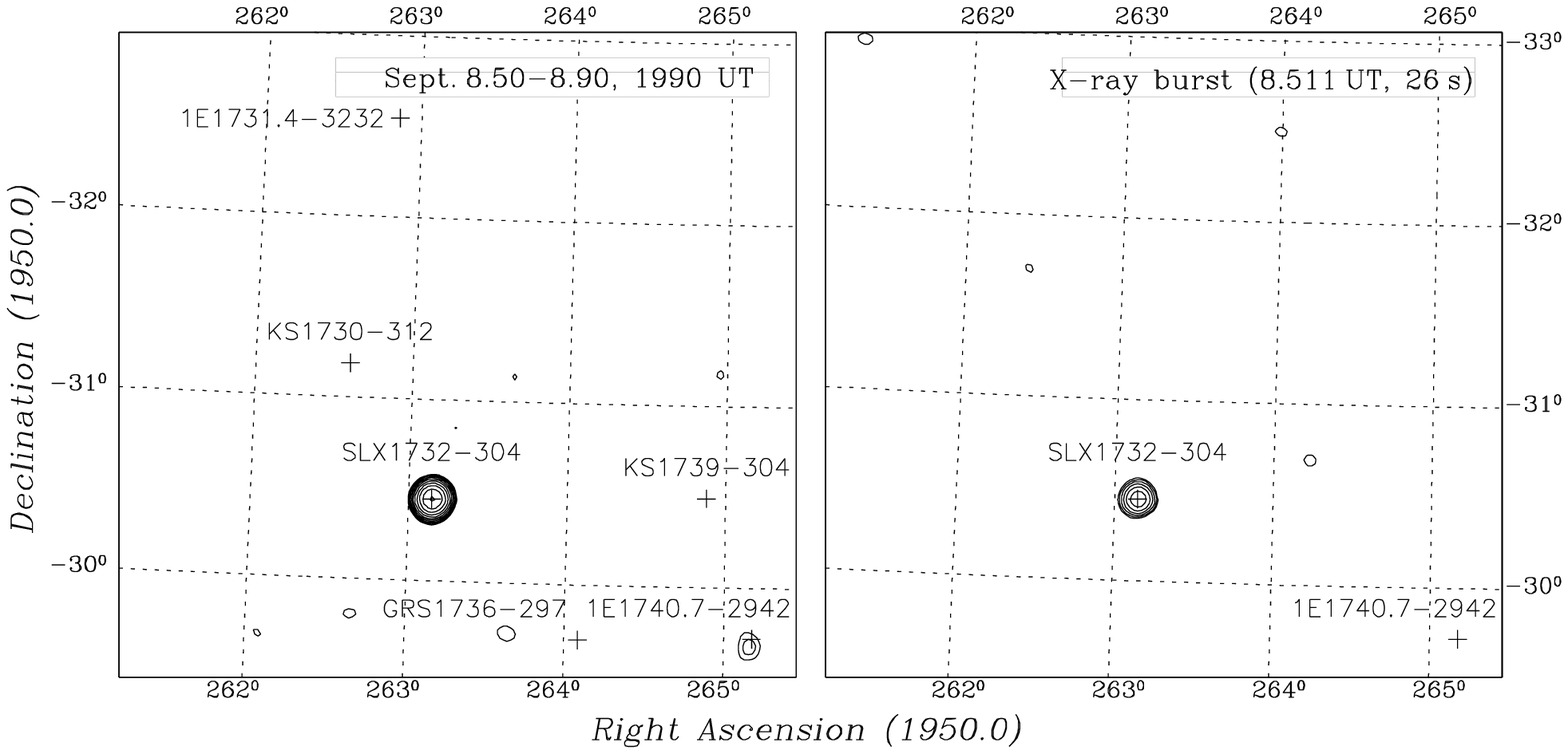}}
\vspace{15pt}
 \caption{\rm ART-P 3-20 keV images of the sky region with SLX 1732-304 (a) 
during the entire observing session on September 8, 1990, and (b) during 
the burst. The 3, 4, 5, or more $\sigma$ confidence regions of the 
source are indicated by contours} 
\end{figure}

\clearpage

\begin{figure}[t] 
\hspace{-2cm}{
\epsfxsize=160mm
\epsffile[20 280 440 590]{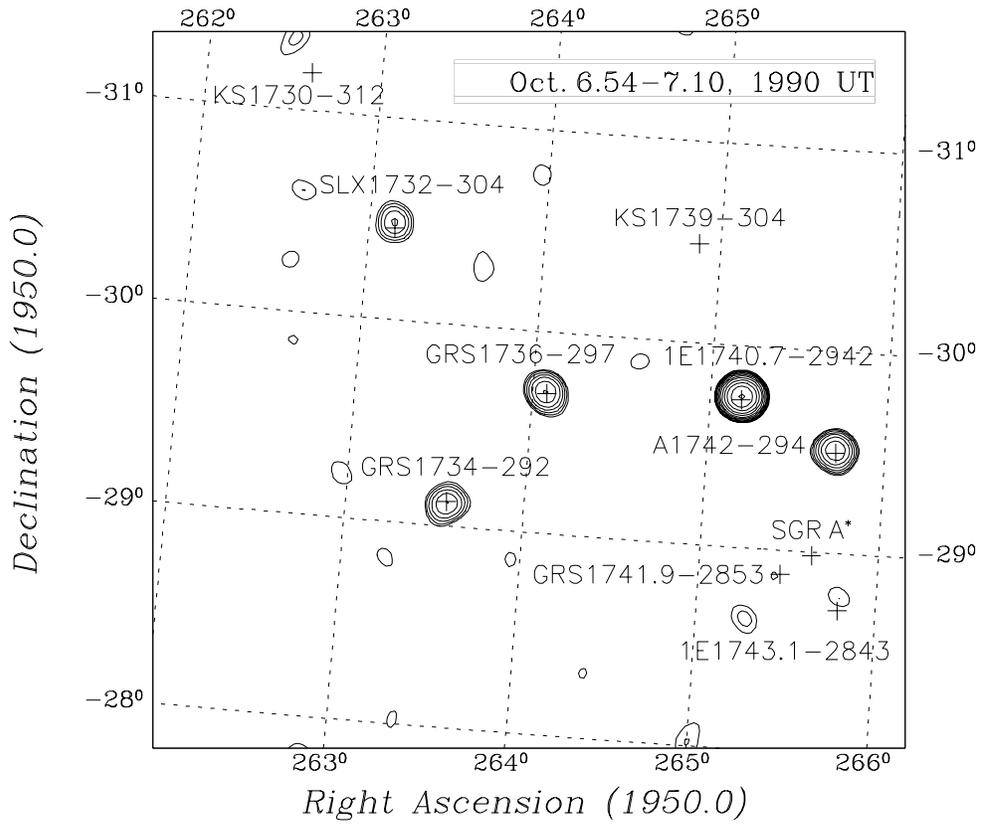}}
\vspace{5pt}
 \caption{\rm ART-P 3-20 keV images of the sky region with SLX 1732-304 during 
the observing session on October 6, 1990. The 3, 4, 5, or more $\sigma$ 
confidence regions of the source are indicated by contours.} 
\end{figure}

\pagebreak

\begin{figure}[t] 
\hspace{0cm}{
\epsfxsize=220mm
\epsffile[18 274 592 718]{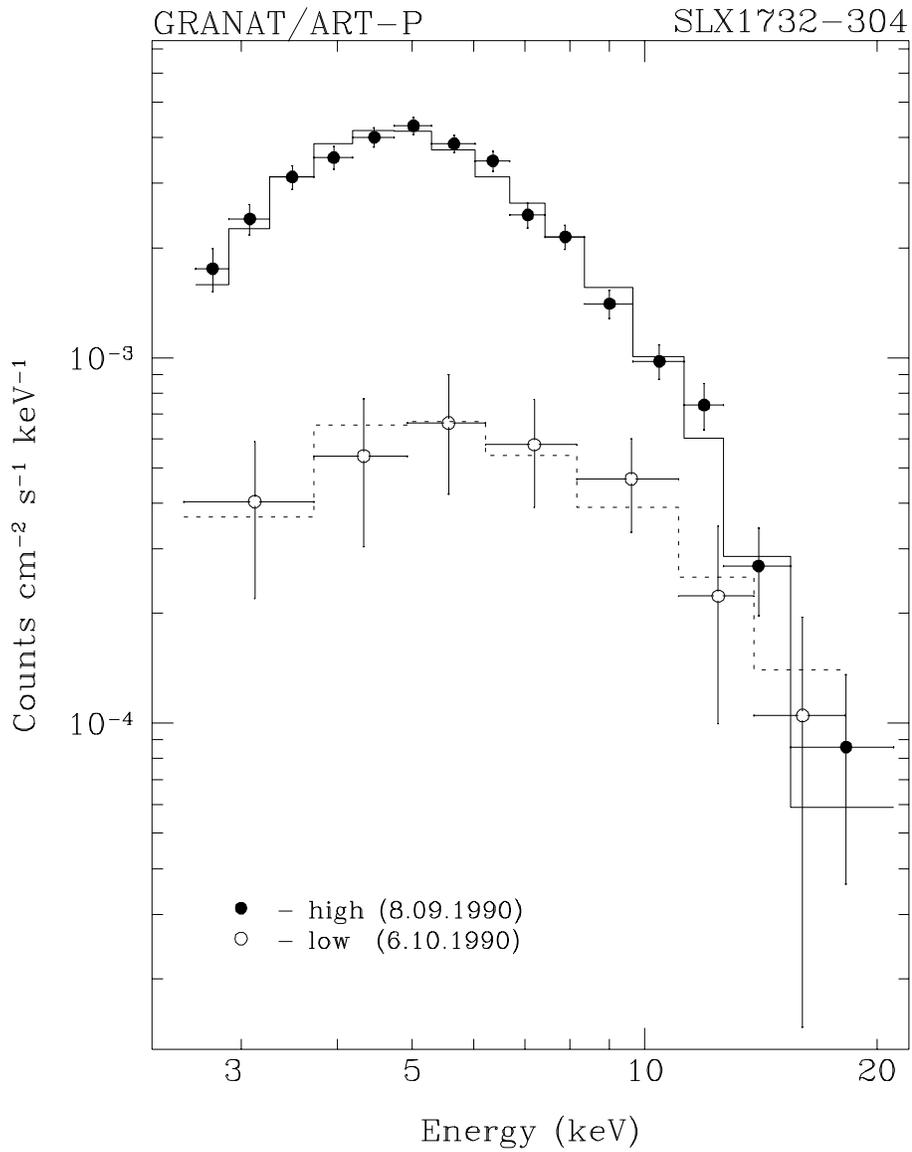}}
\vspace{5pt}
 \caption{\rm Pulse-height spectra for the high (filled circles) and low (open 
circles) states of SLX1732-304 on September 8 and October 6, 1990, 
respectively. The best fits to the spectra are indicated by histograms. 
The errors correspond to one standard deviation.} 
\end{figure}

\pagebreak

\begin{figure}[t] 
\hspace{0cm}{
\epsfxsize=220mm
\epsffile[18 270 592 718]{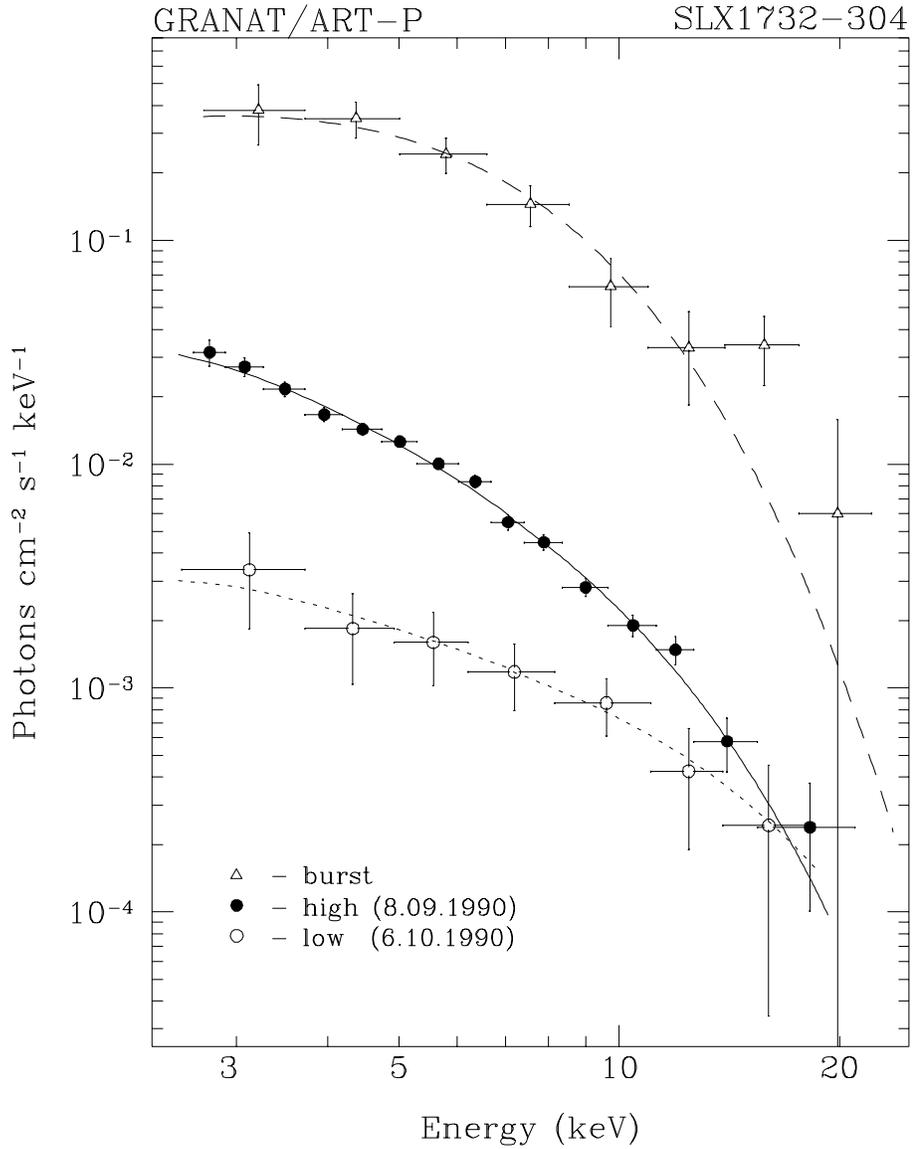}}
\vspace{5pt}
 \caption{\rm Photon spectra for the high (filled circles) and low (open 
circles) states of SLX 1732-304 and for the X-ray burst (triangles). The 
solid, dotted, and dashed lines represent the corresponding model 
spectra. The errors correspond to one standard deviation.} 
\end{figure}

\pagebreak

\begin{figure}[t] 
\hspace{0cm}{
\epsfxsize=210mm
\epsffile[88 460 592 718]{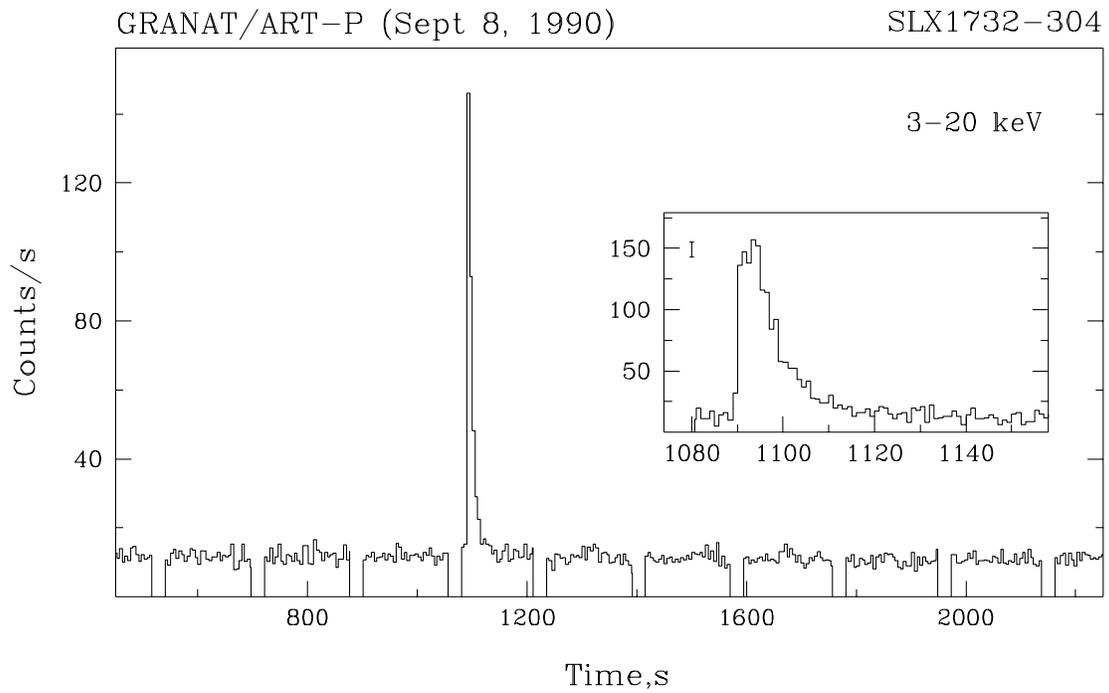}}
\vspace{5pt}
 \caption{\rm Observed light curve of SLX 1732-304 on September 8, 1990. The
insert shows the X-ray burst with the highest time resolution. Time from the
beginning of the session is plotted along the horizontal axis.}  
\end{figure}

\pagebreak

\begin{figure}[t] 
\hspace{0cm}{
\epsfxsize=230mm
\epsffile[68 300 592 718]{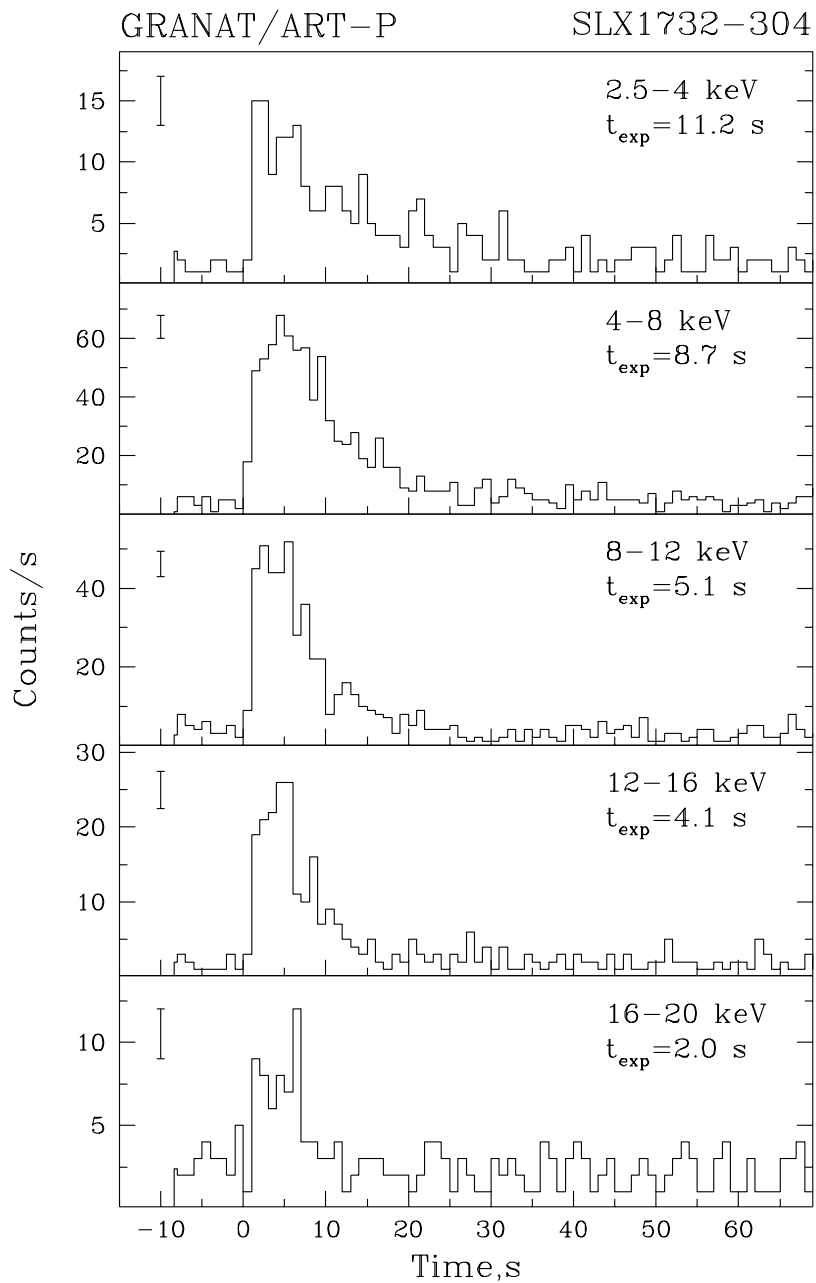}}
\vspace{5pt}
 \caption{\rm Profiles of the X-ray burst in different energy bands
with a time resolution of 1 s; $t_{exp}$ is the $e$-folding decline
time of the source in each band. The errors correspond to one standard
deviation.}   
\end{figure}

\pagebreak

\begin{figure}[t] 
\hspace{0cm}{
\epsfxsize=220mm
\epsffile[18 274 592 718]{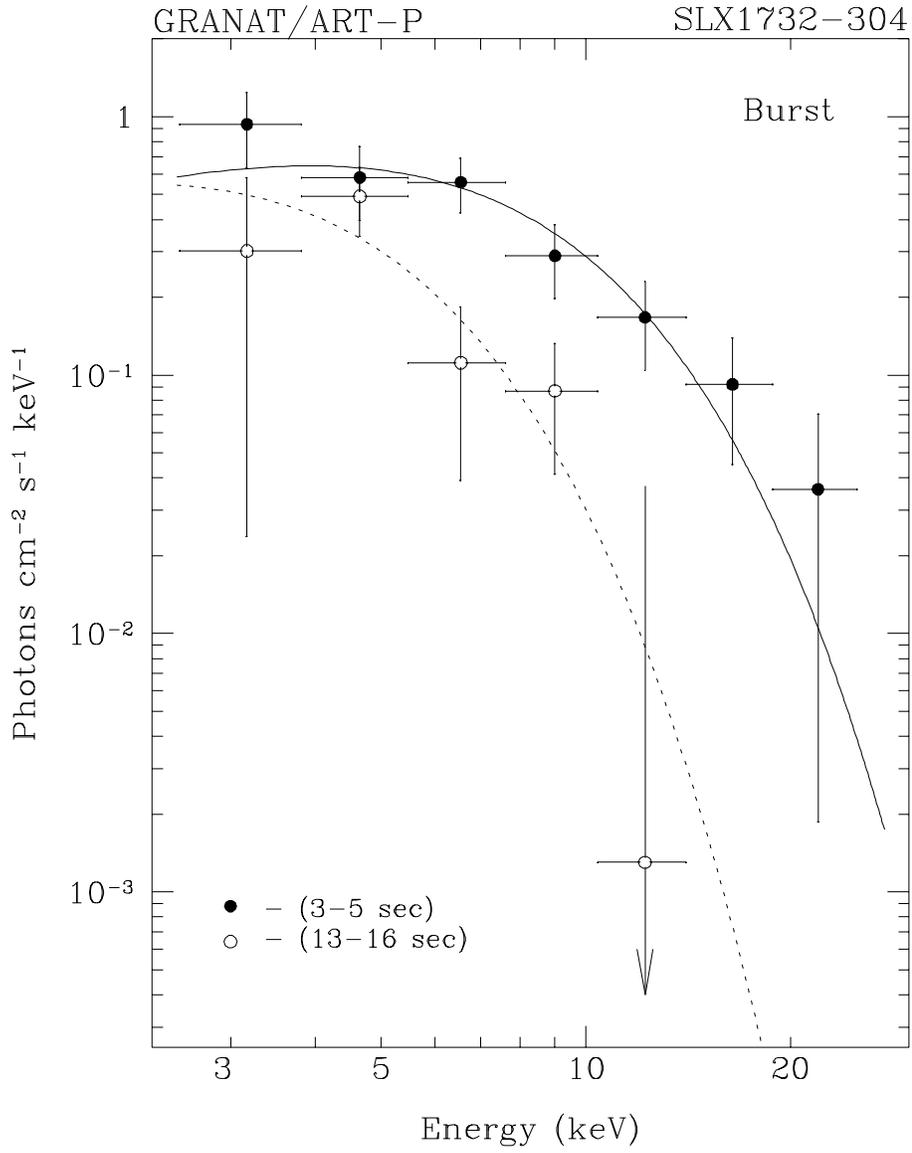}}
\vspace{5pt}
 \caption{\rm Measured photon spectra of SLX1732-304 at the burst peak (filled 
circles) and the end of the cooling stage (open circles). The solid and 
dotted lines represent the corresponding model spectra.}  
\end{figure}

\pagebreak

\begin{figure}[t] 
\hspace{0cm}{
\epsfxsize=220mm
\epsffile[68 320 592 718]{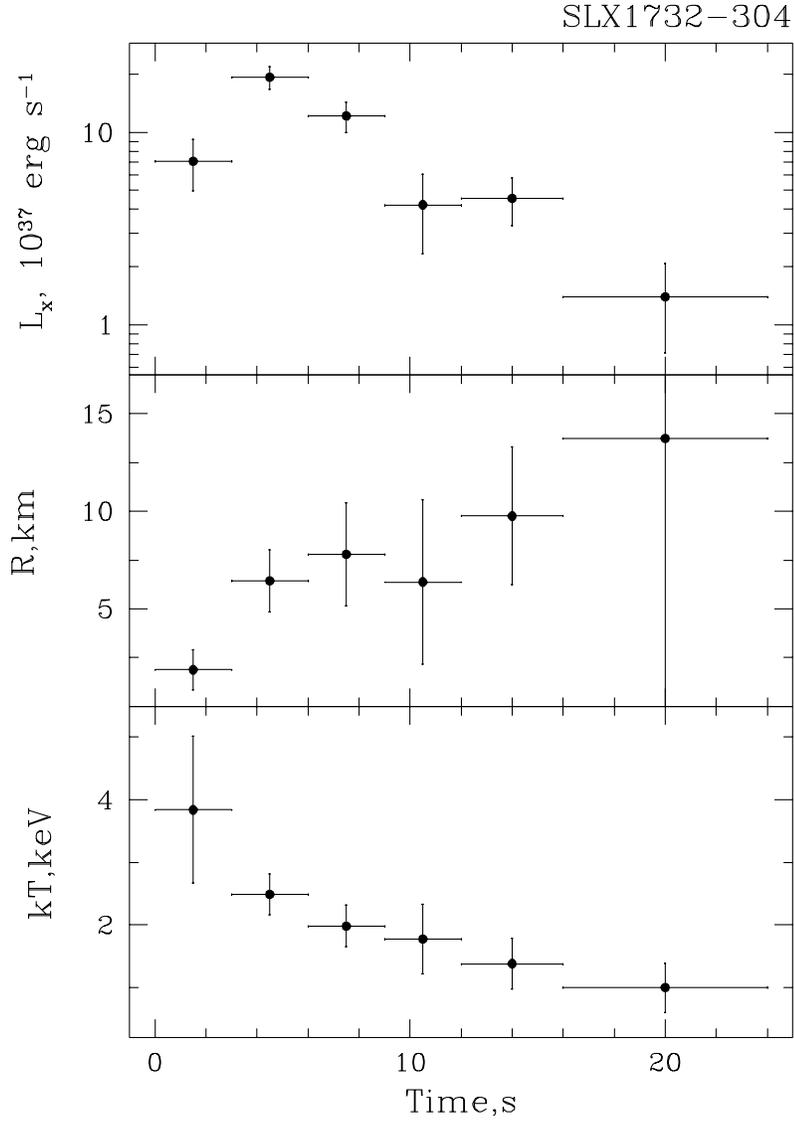}}
\vspace{5pt}
 \caption{\rm Evolution of (a) the source's bolometric luminosity, (b) its 
radius, and (c) the blackbody temperature during the X-ray burst 
detected by ART-P.}  
\end{figure}

\end{document}